# Compressive imaging of transient absorption dynamics on the femtosecond timescale

**ONDŘEJ DENK,**[1] **KAIBO ZHENG,**[2,3] **DONATAS ZIGMANTAS,**[2] **AND KAREL ŽÍDEK**[1,*]

[1]*Regional Centre for Special Optics and Optoelectronic Systems (TOPTEC), Institute of Plasma Physics, Academy of Sciences of the Czech Republic, Za Slovankou 1782/3, 182 00 Prague 8, Czech Republic.*
[2]*Chemical Physics, Lund University, PO Box, Lund SE-22100, Sweden.*
[3]*Department of Chemistry, Technical University of Denmark, DK-2800 Kongens Lyngby, Denmark.*
*\*zidek@ipp.cas.cz*

**Abstract:** Femtosecond spectroscopy is an important tool for tracking rapid photoinduced processes in a variety of materials. To spatially map the processes in a sample would substantially expand the capabilities of the method. This is, however, difficult to achieve due to the necessity to use low-noise detection and to maintain feasible data acquisition time. Here we demonstrate realization of an imaging pump-probe setup, featuring sub-100 fs temporal resolution, by a straightforward modification of a standard pump-probe technique, using a randomly structured probe beam. The structured beam, made by a diffuser, enabled us to computationally reconstruct the maps of transient absorption dynamics based on the concept of compressed sensing. We demonstrate the functionality of the setup in two proof-of-principle experiments, were we achieve spatial resolution of 20 μm. The presented concept provides a feasible route to imaging, using the pump-probe technique and ultrafast spectroscopy in general.

## 1. Introduction

Ultrafast spectroscopy provides us with essential information about processes in many systems of interest, including semiconductor nanostructures, conjugated polymers, or light-harvesting biological complexes. [1–5] By using femtosecond (fs) pulses we can extract dynamics on the corresponding timescale and identify important parameters – for instance, to explore various factors limiting the lifetime of charge carriers in a solar cell. [5] Ultrafast spectroscopy has, therefore, established itself an indispensable tool in material and biological research.

The ability to carry out imaging of measured dynamics elevates the potential of the technique even higher. [6] By employing it one can, for instance, map and identify the "leaking" points of a solar cell, where the carriers rapidly recombine. However, implementation of imaging in ultrafast spectroscopy is a complex issue. We will illustrate this on the pump-probe (P-P) technique – an archetypal technique of ultrafast spectroscopy used to measure the kinetics of transient absorption (TA) or reflectance on the fs timescale. [7]

In the P-P experiment, a strong pump pulse is used to excite photodynamics in a measured sample and the probe pulse, arriving with a certain delay, probes the photoinduced changes in the sample triggered by the pump pulse. The TA signal is measured as a difference in the probe beam intensity for the excited and unexcited sample. Since the measured relative probe change is often very weak, typically in the order of $10^{-3}$ to $10^{-6}$, it is necessary to detect the probe intensity by using a low-noise detector with a high dynamic range, often in combination with a lock-in amplifier, or an equivalent approach. Usually, detection in a P-P setup is carried out by a photodiode array, a cooled line CCD coupled to a spectrograph, or simply a photodiode measuring the total probe intensity.

As far as imaging is concerned, all the commonly used detectors in P-P experiments are single-pixel detectors, which recover information from a single spot in the sample. A straightforward option to carry out imaging is to scan the position of the probe spot in the sample point-by-point. [6,8,9] However, by assuming that the acquisition time of a single datapoint takes mere 3 s (including data acquisition time, sample shift, and the data readout), we reach, for a 100×100 pixel image, the total acquisition time of more than 8 hours. Clearly, measuring several TA maps can last well beyond the feasible time for the experimental setup or the stability of the measured sample.

Another option which has been reported in literature consists in rebuilding the measurement apparatus to detect an image of the probe beam by using an array detector. [10–14] In this case, we gain an image of the sample in a single measurement. However, several drawbacks arise: (i) a suitable detector is an expensive piece of equipment; (ii) spectral information has to be sacrificed; (iii) only sufficiently high TA signal level can be measured.

A solution to the problem, which bypasses both the issues of array sensor and lengthy sample scanning, lies in the theory of compressed sensing (CS) and, more specifically, in compressive imaging. [15,16] We can utilize the fact that information contained in a common image is sparse. This means that in a certain basis, such as Fourier basis, the image can be expressed by using a dataset where the vast majority of components are equal to zero or negligible. In other words, the actual information is contained in a small fraction of the components, which is a fact extensively used in digital image compression (JPEG compression). The CS theory makes it possible to exploit the image sparsity as early as during the image acquisition to minimize the acquisition time and, moreover, to revisit the entire approach to imaging.

One of the most used techniques of CS is denoted as single-pixel camera (SPC). [17] Here, an image of interest is acquired by measuring the total intensity from a random combination of N image pixels. The total intensities measured for a sequence of M different random combinations provide us with enough information to computationally reconstruct the image – even for the cases, where M≪N. We can, therefore, substantially reduce the number of measurements compared to the number of image pixels, yet we can capture the image by using a single-pixel detector. This advantage has been previously employed in various imaging setups ranging from microscopy to time-of-flight imaging. [18–20]

In this article, we demonstrate the construction of an imaging P-P setup based on the concept of SPC. We exploited laser light coherence to generate random patterns (laser speckles), which are used to encode the P-P image. [20] Owing to that, the entire conversion of a standard P-P setup into the imaging one essentially consists in adding a diffuser on a motorized stage into the probe beam, along with a simple reference camera recording the speckle patterns. The use of the patterned probe beam together with a standard P-P detection provide us with the ability to computationally reconstruct an image of a TA signal with unmatched simplicity, while the required acquisition time is several times shorter compared to sample scanning.

The functionality of our P-P imaging setup was verified on several proof-of-principle measurements. In this article we present imaging of a tilted pump beam gradually exciting a homogeneous sample of CdSe quantum dots and imaging of transient dynamics on a structured flat sample.

The presented experiment provides a simple tool for femtosecond ultrafast imaging, which can be used in a number of spectroscopic laboratories. As we will discuss in detail, compared to a sample scanning used on the same setup, the use of CS shortens the data acquisition time, resolves the sample microstructure at a long working distance of tens of centimeters, and makes it possible to drop an arbitrary subset of measurements from the dataset. Moreover, under certain conditions the setup can be used for imaging of probe wavelengths in the infrared or THz spectral regions, where 2D array sensors are not commonly available.

## 2. Experimental setup

We built the P-P imaging setup, which is depicted in Fig. 1, by a minor modification of a classic low-noise P-P experimental setup, which we will hereafter call a 'standard' P-P setup. The standard setup has been used extensively to characterize a variety of samples on the fs and ps timescales. [21,22] We will first briefly describe the standard pump-probe setup and then the changes required to modify the setup into the imaging modality.

*Standard low-noise pump-probe setup*

We used two non-collinear optical parametric amplifiers (NOPAs) to convert seeding laser pulses (Light Conversion, Pharos 1030 nm) to ultrashort visible fs pulses. Two generated beams featuring the central wavelengths of 495 nm and 560 nm (see Fig. 2(A), solid lines) were used as a pump and a probe beam, respectively. The probe beam was directed through a time delay line to adjust the P-P delay. The laser repetition rate was set to 20 kHz in order to ensure that the testing samples return to their ground state before the arrival of the following pulse. The pump beam was modulated using an optical chopper.

The probe beam intensity at a sample was set for both the imaging and the standard setups to 1 nJ/pulse. The attained pump pulse energy of 40 nJ/pulse at a sample was fully used in the imaging experiment, because the beam covered a relatively large area of the sample (0.5-2 mm2, exc. energy density of 2-8 μJ/cm2). In the standard P-P setup the pump beam was attenuated to generate a TA signal level equal to the imaging setup.

The pump and probe pulses were compressed using a combination of chirp mirrors and prism compressors into ultrashort pulses with the FWHM of 21 fs and 22 fs, respectively. The pulse length was determined based on the second harmonic autocorrelation of the pulses – see Fig. 2(B). Analogously, the response function of the standard setup was acquired using a sum-frequency generation of the pump and probe pulses (FWHM 33 fs).

The pump and the probe pulses were focused onto a sample using spherical mirrors and the transmitted pump beam intensity was recorded by using a Si amplified photodiode (New Focus). The signal from the photodiode was digitized by using a data acquisition card and processed by the means of in-house lock-in detection analysis algorithm.

*Conversion to the imaging setup*

In the imaging setup in Fig. 1, the pump beam followed the standard P-P scheme. However, the sample was placed behind the focal plane of the focusing mirror to gain an excitation spot large enough to cover the area of interest for imaging. The used pump spot was elliptically shaped with approximately 1×0.5 mm dimensions.

The probe beam was focused by a spherical mirror onto a standard diffuser, giving rise to many laser wavefronts with random phases. The wavefronts form by their mutual interference a random speckle pattern which can be in the far-field regime considered as a spherical wave with a modulated intensity in various directions. Therefore, it was possible to collimate the speckle pattern into a beam by using an off-axis parabolic mirror. The mean size of the speckles in the collimated beam, i.e., the pattern fineness, is an important factor for the setup spatial resolution, as we showed in our previous work. [20] The fineness was determined by the size of the probe spot on the diffuser. The generating diffuser was placed on a motorized stage to attain a set of different speckle patterns by shifting the diffuser to different positions.

An iris diaphragm was used to cut out the central part of the collimated speckle beam, which was reflected via a spherical mirror onto the sample. The sample was placed well before the focal plane of the focusing spherical mirror. Close to the sample, the patterned probe beam was partly reflected by a thin beam splitter (BK7 slide with an anti-reflective coating on one side) and a reference probe image was acquired by a reference camera. The camera was placed to match the sample – beam-splitter distance. However, since the sample was not in the beam waist proximity (Rayleigh range), the reference camera position could be set roughly without losing a correlation between the sample and the reference speckle pattern. [23] The reference image was attained by a CMOS chip (IDS uEye) with a neutral density filter reducing the beam intensity.

To summarize, all that is required for the transformation of the standard P-P setup into the imaging one is the generation of a patterned probe beam using a diffuser on a motorized stage, and the use of a reference camera recording a reflected probe beam reference.

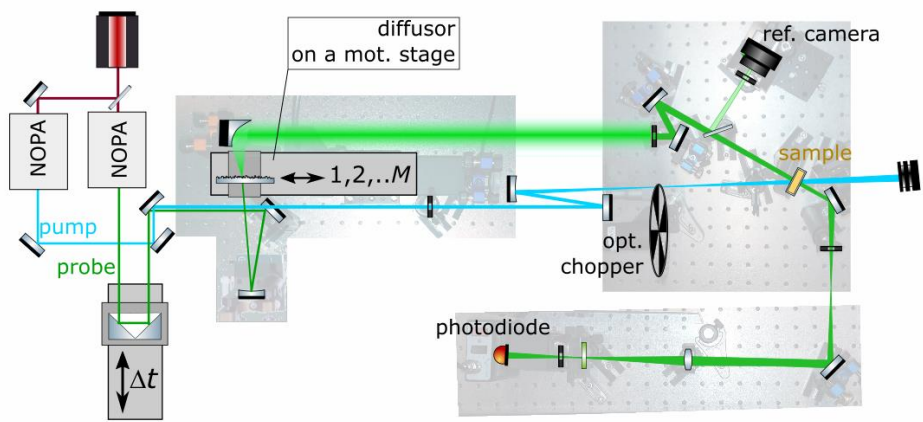

Fig. 1. Setup scheme used to carry out pump-probe imaging with a tilted pump beam – see text for details.

*Testing samples*

The experimental setup was tested by using a sample of colloidal CdSe quantum dots (QDs) with the lowest excitonic absorption maximum located at 560 nm (see Fig. 2A). [24] This corresponds to the mean size of the QDs of about 3.4 nm. A solution of the QDs was placed in a 0.5 mm thick cuvette, featuring the optical density of 0.3-0.4 in the studied spectral range. The QDs had been previously extensively studied with respect to their optical and structural properties. [25–27]

The second testing sample was a structured sample created by drop-casting a solution of QDs and Rhodamine 6G on a microscopic slide. After drying, the solutions created islands of agglomerated QDs and Rhodamine 6G molecules. In the case of the QDs, the spectral features do not change with the drop casting, whereas the Rhodamine 6G molecules encounter a shift in their resonances.

## 3. Results and discussion

### 3.1 Standard pump-probe setup

By using a standard pump-probe setup we were able to track TA kinetics in the solution of QDs and at several spots of the structured sample – see Fig. 2(C). The probe spectrum was chosen to match the region of the lowest excitonic state of the used QDs. Since the exact origin of the dynamics is not important for this work, we discuss it only briefly.

In the case of the colloidal sample (red line), the TA signal is dominated by a strong bleach of the lowest absorption band of the QDs. The kinetic follows an expected behavior, where we can observe a very rapid relaxation of carriers into the lowest excited state (TA signal onset) followed by a long-lived TA signal in the single-exited QDs decaying due to the electron-hole recombination. The excitation fluence is very low and thus we can neglect any effects arising from the multiply-excited QDs. [27]

In the structured sample (black lines), in addition to the previous effects, we observe a rapid onset and decay in the TA signal due to surface traps induced on the surface of the QDs and a range of lifetimes in a variety of Rhodamine 6G agglomerates. [28,29]

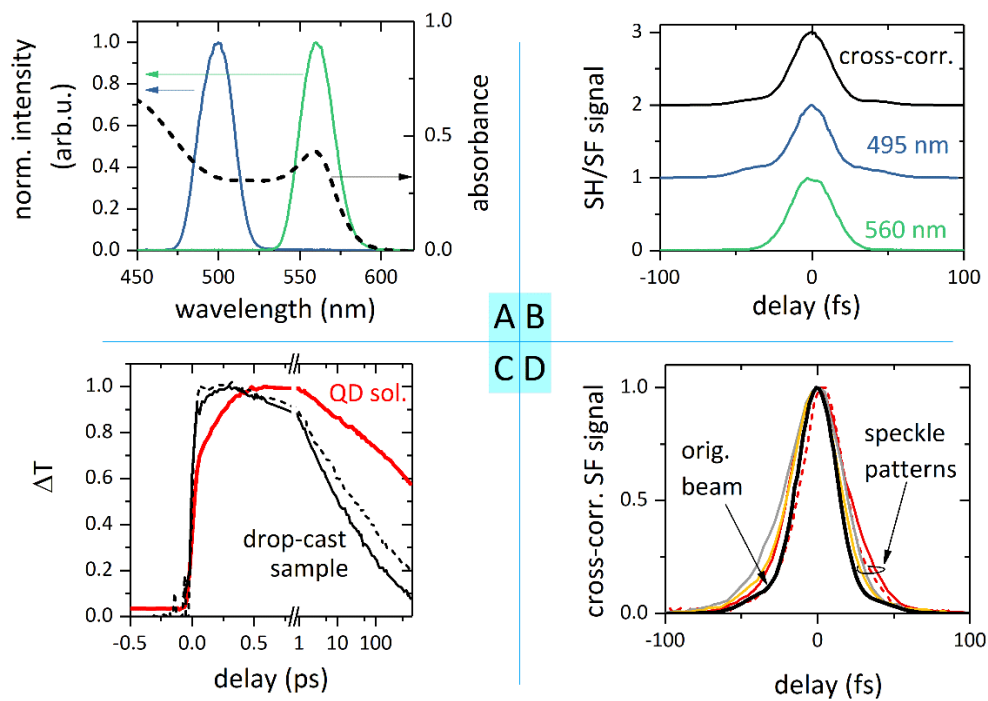

Fig. 2. A) Spectra of the pump (blue) and probe (green) pulses. Absorbance of the measured quantum dots (dotted line). B) Autocorrelation of the pump and probe beams together with the cross-correlation signal. C) TA kinetics recorded by a standard PP setup for a colloidal sample (red line) and structured sample (black lines). D) Cross-correlation between the pump beam and speckle patterned probe beam for various diffuser positions (colors) and selected regions (red lines) compared to the original cross-correlation (black line).

## 3.2 Patterned Speckle Beam for a Single-pixel camera

The imaging is based on the method denoted as single-pixel camera (SPC). [17] We point the reader to introductory articles on this topic, which elucidate this method and CS in general. [15,30] In the following, we provide a brief overview.

Using the SPC concept we need to illuminate the measured sample by a series of known uncorrelated random patterns. In our case, we used random speckle patterns which are created when coherent laser light is transmitted through a diffuser. This is due to interference of many wavefronts with random phases. By shifting the diffuser by sufficiently long steps, we attain a series of uncorrelated speckle patterns, as we showed in our previous work. [20]

The random modification of a phase on the diffuser, however, also implies that the transmitted pulse is randomly delayed within the speckle pattern. Due to this effect, we observe that the patterned probe pulses are stretched in time. This was determined by measuring the cross-correlation between the pump and the patterned probe beam via sum-frequency generation. The cross-correlation, as a result of variable delay within the patterned probe pulse, varies with changing diffuser position (different colors in Fig. 2(D)) as well as between various spots in one pattern (red solid and dashed lines in Fig. 2(D)). The variation between pulses limits the temporal resolution of the imaging setup. Yet, the setup response function does not exceed the FWHM of 50 fs, which is sufficient for many processes of interest studied with ultrafast spectroscopy.

We will now explain the basic idea behind the P-P imaging using a patterned probe. When a patterned probe beam with the energy density of $I(x, y)$ is transmitted through a sample, the recorded value is different in the case where the sample is in its ground state ($I^{gs}$) and where the sample is excited ($I^{exc}$). The value is affected by a photoinduced change in transmittance $\Delta T(x, y)$, i.e., the TA signal. The corresponding measured lock-in signal $s(x, y)$ is:

$$s(x, y) = I^{exc}(x, y) - I^{gs}(x, y) \approx \Delta T(x, y) I_{pr}(x, y). \quad (1)$$

We employed here the fact that $\Delta T \ll 1$ since a small fraction of a sample is typically excited. By detecting the total probe intensity (i.e., by integrating over the whole probe spot) we get:

$$s_{TOT} = \iint s(x, y)\, dxdy = \iint \Delta T(x, y) I_{pr}(x, y)\, dxdy = \sum_{i,j} \Delta \mathcal{T}_{i,j} \mathcal{J}_{i,j}. \quad (2)$$

In the last term, we discretized the TA signal and the probe intensity into the form of matrices $\Delta \mathcal{T}$ and $\mathcal{J}$. The physical meaning of the TA signal remains unchanged; the discretized probe intensity $\mathcal{J}$ now denotes the probe energy in one pixel of the image. For the sake of simplicity, we can convert 2D matrices $\Delta \mathcal{T}$ and $\mathcal{J}$ into two vectors $\Delta \tilde{T}$ and $\tilde{I}$ sized $N \times 1$ and $1 \times N$, respectively, where $N$ denotes the number of dataset elements (i.e., number of image pixels). The Eq. (2) can be then rewritten by using a matrix multiplication as

$$s_{TOT} = \Delta \tilde{T} \times \tilde{I}. \quad (3)$$

When measuring the lock-in signal for $M$ different probe patterns, we can create a matrix $\mathcal{A}$ where each line of the matrix will be formed by a single probe pattern $\tilde{I}$. The lock-in signal values $s_{TOT}$ can be placed in a $1 \times M$ vector $\tilde{S}$ and the vector is related to the TA signal as:

$$\tilde{S} = \mathcal{A} \times \Delta \tilde{T}. \quad (4)$$

Our goal is to extract the TA signal image $\Delta \tilde{T}$ while both vector $\tilde{S}$ and matrix $\mathcal{A}$ are known, since we acquire the speckle pattern reference and the lock-in signal for each speckle pattern. Yet, it is possible to obtain $\Delta \tilde{T}$ directly only if $\mathcal{A}$ is an invertible matrix. This cannot be satisfied for $M < N$.

Nevertheless, CS makes is possible to exactly recover the $\Delta \tilde{T}$ even for a highly underdetermined system ($M \ll N$). This is conditioned by three prerequisites: firstly, the $\Delta \tilde{T}$ vector has to be sparse or sparse-like in certain basis – as stated before. This condition is satisfied for a vast majority of images. Secondly, the measured data needs to be encoded by random uncorrelated patterns. Finally, it is required to mathematically quantify an expected characteristics of the reconstructed image with a so-called regularizing term. Here, the CS theory benefits from extensive research in the field of image processing. One of the most commonly used properties is the total variation $D$ of an image, which is an $l^2$-norm of the difference between the neighboring pixels in an image:

$$D(\Delta \mathcal{T}_{i,j}) = \sqrt{\sum_{i,j}(\Delta \mathcal{T}_{i,j} - \Delta \mathcal{T}_{i-1,j})^2 + \sum_{i,j}(\Delta \mathcal{T}_{i,j} - \Delta \mathcal{T}_{i,j-1})^2}. \quad (5)$$

The total variation of an image is low for the cases where the image contains a small number of transitions between the dark and bright pixels, which is the case for the vast majority of images. On the contrary, noisy random images feature a high value of total variation.

We can combine our knowledge of the relation between the measured data and the TA map in Eq. (4) and, consequently, we can apply the condition that the resulting TA map is expected to feature a low value of its total variation, equation (5). The image reconstruction can, therefore, be carried out by minimizing both the factors:

$$\Delta \tilde{T}_{rec} = arg\min_{\Delta \tilde{T} \in R^N} \left\{ D(\Delta \tilde{T}) + \frac{\mu}{2} \left\| \tilde{S} - \mathcal{A} \times \Delta \tilde{T} \right\|_2^2 \right\}. \quad (6)$$

The weight factor $\mu$ is added either to promote the importance of the measured dataset for a high $\mu$ value, or to promote the minimization of the total variation when the $\mu$ value is low. We will closely discuss the factor value and its effect in the text below. Nevertheless, Eq. (6) represents a standard problem in CS-based imaging, which can be solved by using the TVAL3 algorithm. [31]

## 3.3 Proof-of-principle measurements

As a proof-of-principle experiment, we carried out imaging of a fully homogeneous sample (solution of CdSe QDs) excited by a tilted pump beam – see Fig. 3(A). The angle between the pump and the probe beam was intentionally set to the high value of 35 deg, which was determined by the relative position of the measurement spot on the sample and the pump and probe beam spots on the preceding mirrors. The sample was placed perpendicularly to the probe beam. Due to the tilt, the sample became gradually excited from the right-hand side as the P-P delay increased. Since the observed bleach in the excited QDs is long-lived and does not significantly decay during the measurement window, we observe, in fact, a gradual onset of the TA signal through the sample. We extracted TA maps which are in a perfect agreement with the expected behavior – see Fig. 3(B).

In this experiment, the size of the probe beam on the sample was set large (2.5 mm in diameter), exceeding the pump beam spot size. This means that the outer parts of image remain unexcited without any TA signal, and the central excited part follows the shape of the pump beam. Indeed, the TA map for the completely excited sample in Fig. 3(B), bottom image, matches the pump spot on the sample in Fig. 3(C). Note, that the pump spot is more elongated in panel C along the x-axis due to the tilt between the pump beam and the sample.

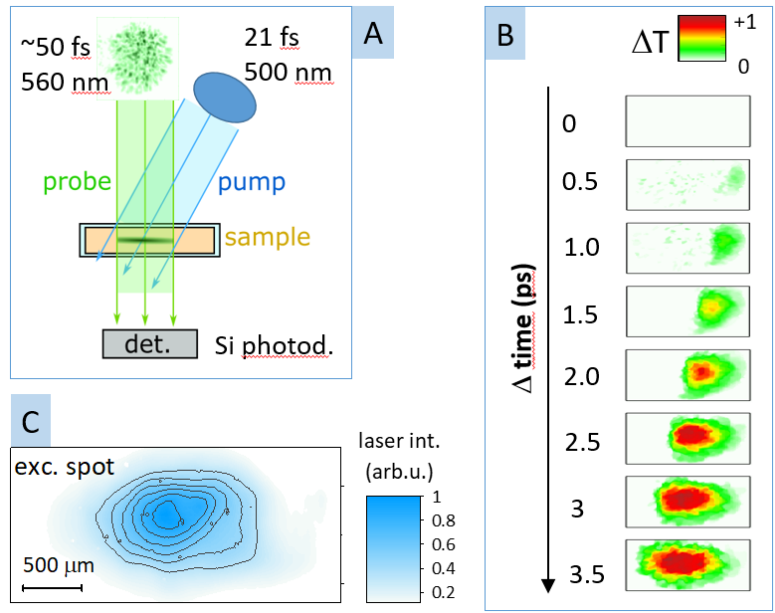

Fig. 3. Pump-probe imaging of a tilted pump beam ($N = 128 \times 128 = 16384$ pixels, $M = 1600$, $M/N_{ph} = 0.27$). A) Scheme of the measurement. B) TA signal maps for pump-probe delay 0-3.5 ps. C) Map of the excitation spot intensity.

The attained dataset of 1600 speckle patterns and the corresponding total TA signal for a set of P-P delays make it possible to reconstruct the kinetics of the TA signal at each spot. We can, therefore, extract dynamics for different spots along the x-axis in the center of the image – see Fig. 4(A), solid lines. The curves show, as expected, a gradual excitation of the sample, often referred to as a "travelling wave" excitation. We have globally fitted the data using a step-function convoluted with a Gaussian response function (dotted lines), where the response function width was fitted as a common value for all curves. This enables us to extract the zero P-P delays for each curve, see Fig. 4(B), which lie, within the fit precision, on a line corresponding to the change of 36.9 fs/pixel – dotted line in Fig. 4(B). The only exception is the last point, where the onset was not fully captured in the measured time range. Therefore, the zero delay is likely to be underestimated.

Since the dimensions of an image pixel in Fig. 3(B) can be calculated from the reference camera snapshots and corresponded to the size of 19 μm, we can derive that the onset of the TA signal was travelling the distance of 1 μm along the x axis in 1.9 fs. From this value and the speed of light we can derive the tilt of the pump beam as $\sin\theta = (1.9\frac{fs}{\mu m})/(0.3\frac{\mu m}{fs})$, which implies that the angle between the P-P beams in air was 35 deg. This is in a perfect agreement with the set angle. Note that the pump beam is refracted in a cuvette and QD solution. However, since it propagates there at a lower speed due to a higher refractive index, the effects cancel out and the TA onset travelling speed remains the same.

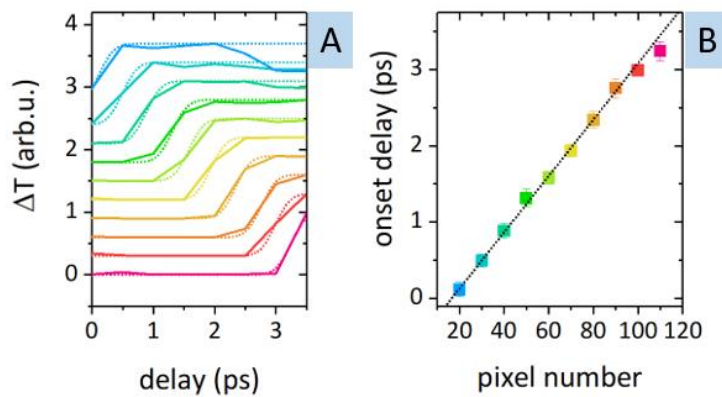

Fig. 4. Gradual sample excitation kinetics. A) Normalized TA kinetics extracted for a single line (64) with varying pixel x coordinates (blue-red: 20, 30, 40, 50, 60, 70, 80, 90, 100, 110) – solid lines, global fit of the kinetics with a step function convoluted with a Gaussian function ($\sigma = 140$ fs) – dotted lines. B) Fitted onset delay for each of the TA kinetics (squares) fitted with a linear function (dotted line, 36.9 fs/pixel image).

In the experiment presented above we were able to verify the setup functionality and capture TA images which are in full agreement with the expected physical behavior. Thus, we turned to the measurement of the structured drop-cast sample on a microscopic slide described above. Here, the probe spot size was decreased to be fully covered by the pump spot, i.e., the retrieved image was not distorted by the pump beam shape. At the same time, the tilt between the pump and the probe beams was set to approximately 10 degrees to avoid the effect of the gradual sample excitation. Due to a lower sample optical density the TA signal was ~6 times lower as compared to the previous measurement.

The structured sample consists of overlapping islands of aggregated QDs and Rhodamine 6G molecules, which, despite not being pronounced under ambient illumination in Fig.5(A), can be clearly distinguished by their photoluminescence, as seen in Fig. 5(B) (excited by 488 nm cw laser, 1 W/cm$^2$). The white circle in Fig. 5(B) denotes the probe spot coverage.

From 800 measurements we were able to retrieve TA maps for a series of pump-probe delays – see Fig. 5(C). Due to the lower signal level and a relatively low number of the measured data points, the reconstructed maps were considerably noisier compared to the previous experiment. In the case of the drop-cast sample, the signal to noise ratio reached the value of around 1.5. Yet, in spite of a relatively low signal level and a 7 times reduced number of measurements as compared to the sample scan (discussed later in the text), we were able to retrieve the TA maps in Fig 5(C), featuring patterns which match the structure observed via photoluminescence.

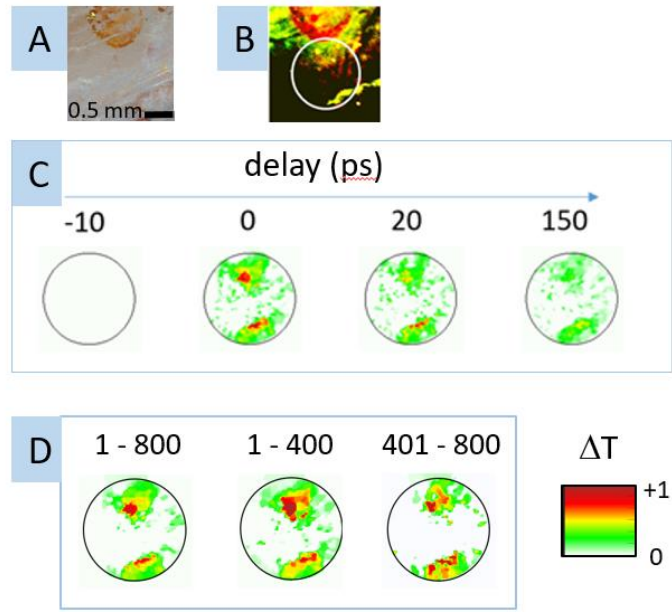

Fig. 5. Imaging of a structured sample – drop-cast Rhodamine 6G and CdSe QDs ($N = 128\times128 = 16384$ pixels, $M = 800$, $M/N_{ph}= 0.14$). A) Sample under microscope under an ambient light. B) Photoluminescence of the sample under 488 nm cw laser excitation. C) Calculated TA maps for a complete dataset ($M = 800$) at given delays. D) Comparison of TA maps $\Delta T(0\ \text{ps})$ for a complete dataset ($i$:1-800; $M = 800$) and two independent subsets ($i$:1-400; $i$:401-800; $M = 400$).

Unlike in standard imaging, the noise level in the resulting image can be reduced by decreasing the μ factor in equation (5) and other parameters of the TVAL3 algorithm. In order to illustrate this fact, we have reconstructed the measured data on the structured sample for several sets of parameters – see Fig. 6 (i) to (v). The upper line depicts the peak TA signal, whereas the lower line is a ratio of TA signals $\Delta T(20\ \text{ps})/\Delta T(0\ \text{ps})$ reconstructed for the same parameters.

First, from the standard P-P measurement we know that the TA signal in the studied samples was only positive. We can, therefore, choose either to restrict the reconstruction to the positive values of $\Delta T$ (Fig. 6, i-iii) or to solve the general case (iv-v). For both options we can vary the parameter μ in equation (6). For a lower μ value (i, iv), we enhance the importance of the total variation over the agreement with the measured TA data. On the contrary, for a high μ value (iii, v) we promote reconstruction which agrees with the measured TA data. This, of course, causes the image to be significantly noisier. In the panel (ii) the μ value is set in between the extreme cases.

We observe that the different reconstruction parameters provide us, in general, with the same picture of the sample, i.e., two prominent parts divided by an area with a lower TA signal. The image of the TA signal is in good agreement with the measured photoluminescence image showing a pronounced region of QDs, especially for the high μ values. In all cases (i-v), we also observe that the sample in the upper part decays more rapidly compared to the lower part. However, the exact shape and size of the reconstructed islands clearly depend on the value of μ.

This raises the essential question regarding the correct selection of the μ value. For instance, when the question of interest concerns the mean dimensions of the islands, the reconstruction parameters will have a huge effect on the result. In general, it is beneficial to use a high μ value promoting the weight of the measured data in the TA map reconstruction. This, on the other hand, gives rise to a number of small features in the image, and it is unclear whether they carry additional information about the sample or are consequence of the noise. To distinguish between the two cases, we can utilize one of the important properties of SPC, namely that each of the measurements provides information about the entire image. We can, therefore, reconstruct a TA map from an arbitrarily selected subset of measurements.

In our case, we split the 800 measurement datasets into two parts and reconstructed TA maps for the two partial datasets – see Fig. 5D. We observe analogous features in both cases, which were taken for entirely different speckle patterns and probe intensities. The resulting images are noisier because the lower number of patterns reduces the reconstruction precision for the noisy datasets. Nevertheless, it is possible to compare the two maps and identify regions common in both reconstructions, and thus identify sample features as opposed to the ones appearing form noise.

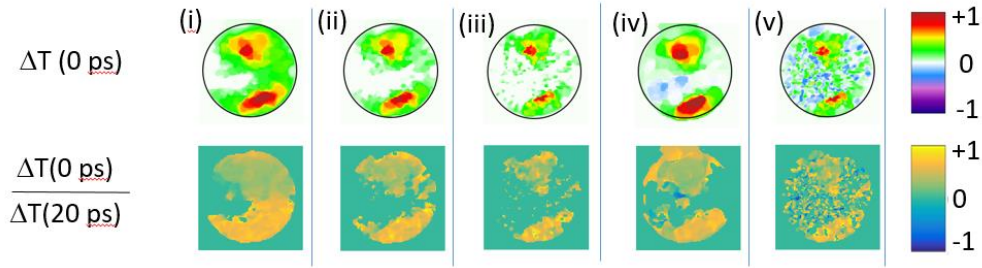

Fig. 6. Imaging of the structured sample – the effect of the reconstruction parameter μ. Upper line: TA maps for a zero delay. Lower line: maps of $\Delta T(20\ ps)/\Delta T(0\ ps)$. i) μ=90, $\Delta T \geq 0$; ii) μ=180, $\Delta T \geq 0$; iii) μ=510, $\Delta T \geq 0$; iv) μ=90, $\Delta T \in \mathbb{R}$; v) μ=510, $\Delta T \in \mathbb{R}$.

*3.5 Advantages and limitations of the method*

In comparison to pixel-by-pixel scanning, the SPC-based approach has several advantages: (i) decreased number of measurements; (ii) attainable image resolution at a long working distance; (iii) possibility to discard arbitrary data points. We will address each of the advantages in turn.

The number of measurements *M* needed to reasonably reconstruct an image encoded by laser speckles is given both by the measurement noise level and the physical resolution of the image. The physical resolution can be different from the number of pixels of the reconstructed image. The number of pixels is only limited by the resolution of the reference camera capturing the reference speckle images. In our measurements, the speckle patterns covered approx. 450×450 pixels and, in principle, we could solve equation (6) by using the speckle patterns in their full resolution and reach the number of pixels of 450×450 = 202500 pixels.

However, as we showed in our previous work, the actual physical resolution of the speckle-based imaging is limited to one half of the mean size of the speckles in the used speckle patterns. [20] The mean size can be determined based on the autocorrelation of the speckle patterns. For the 450×450 speckle images, the mean speckle size was 9.7 pixels and 11.5 pixels in the horizontal and vertical sense, respectively.  This value stayed constant for all measurements. Hence, the physical resolution of our images reached 82×72 pixels (N = 5900). Note that, for practical reasons, we performed all data processing by rescaling the raw speckle patterns to 128×128 pixels prior to the TVAL3 reconstruction.

The images were reconstructed using 800-1600 measurement data points, which correspond to 14-28 % of the number of pixels for the physical resolution. In other words, the use of CS helps to reduce the number of measurements, compared to sample scanning, by the factor of 5-7. For a single measurement requiring about 5 s/data point (4 s signal averaging, 1 s for stage and delay adjustments), we achieve the measurement time of a single image of 1-2 hours for the CS as opposed to 8 hours for the pixel scan. Such difference is very important from the point of view of laser system stability, sample photostability, etc.

Since we attained the resolution of 82×72 on a probe spot of 1340 μm in diameter, we can resolve details down to approx. 20 μm. For the pixel scan, the same resolution can be accomplished by focusing the probe beam onto a comparable spot. Practically, reaching such resolution would be conditioned by using microscopy optics implying a short working distance. In our case, we used standard spherical mirrors with a focal length of 10-25 cm. The experiment can be, therefore, easily used in combination with a cryostat, a magnet or similar devices, unlike a microscopy apparatus.

As stated above, we can reconstruct the image from an arbitrary subset of the measured data. We used this in our measurements to discard the speckle patterns, where the reference camera was saturated and could not provide full information about the pattern. Another possible application consists, for instance, in dealing with measurements where a random source of noise disturbs some data points. In the pixel scanning method the noise would corrupt certain areas of the image. In our method, such data points can be simply omitted.

A principal limitation of the presented imaging method is the fact that we have to have a basic preliminary knowledge of the measured sample to apply the method correctly. Generally, the used image reconstruction is based on the assumption that the desired image features a low value of total variation. Although this is true for the majority of cases, samples with a pronounced structure on a scale comparable to the imaging resolution will pose a problem. This fact can be, on the other hand, verified by comparing reconstruction from multiple independent subsets of the measured data.

The method also features limited precision for the exact lifetime evaluation. Even though the reconstruction provides us with a good comparison of the signal strength for a given delay, decay curves might be distorted due to minimization of total variation. An example can be the decay curves in Fig. 4(A), where the initial signal on the right hand side (blue-green curves) is promoted. For this reason, the TA decay seems to vary from the blue to the yellow curves. In a homogenous solution of single-excited QDs, this is very likely to be an artefact of the mathematical reconstruction.

Finally, we would like to point out one very useful feature of the presented P-P imaging. In the cases where the TA signal strength linearly scales with the excitation intensity – and this applies in many cases where a low excitation level is used — we can derive Eq. (4) in an analogous way, where we use a homogeneous probe beam and a patterned pump beam. This means that the TA imaging can be, in such case, carried out by using a patterned pump beam, instead of the probe beam. Such approach is very appealing in the cases where the pump beam is in the visible/near-IR range whereas the probe beam can be in any spectral range (IR, THz, …). The pump beam patterned can be easily captured by a reference camera and we still retain the advantage of measuring the probe beam with a single-pixel detector. Such experiment can, for instance, open the path for ultrafast THz imaging.

## 4. Conclusions

We presented a concept which offers a straightforward access to imaging in the field of ultrafast spectroscopy. The presented pump-probe setup is able to map transient absorption signal with a sub-100-fs resolution by using a fraction of measurements compared to a point-by-point sample scanning. This is achieved by employing a patterned probe beam, generated by a diffuser, to randomly encode the transient absorption signal. We have demonstrated the functionality of the setup on two proof-of-principle measurements. The experiment was able to reproduce a gradual excitation of a sample from a tilted pump beam and an inhomogeneous TA signal from a structured sample.

Owing to its simplicity, the presented approach can become simple means of imaging in ultrafast spectroscopy for a broad range of laboratories. The conversion from a standard time-resolved spectroscopy setup is straightforward and although the presented article focuses solely on the pump-probe setup, an analogous experiment can be carried out for any technique where the resulting signal is proportional to the probe beam or a local oscillator in heterodyne detection, e.g., the transient grating technique or certain types of two-dimensional spectroscopy.

Moreover, for the pump-probe signal, which is proportional to the pump intensity, it is possible to create an analogous experiment, where a patterned pump beam is used to encode the image. In this case, the imaging can be carried out for a broad spectral range of probe beams, including THz.

## Funding

Laserlab-Europe EU-H2020 (grant agreement 654148); Grant Agency of the Czech Republic (Project 17-26284Y); Independent Research Fund Denmark-Sapere Aude starting grant (No. 7026-00037A)

## Acknowledgment

We thank Lukas Wittenbecher (Lund University) for his help during the experimental work and Ramūnas Augulis for the lock-in signal analysis.